\documentclass{PoS}

\title{FINAL STATE DISTORTIONS IN TWO-PARTICLE CORRELATIONS}

\ShortTitle{FINAL STATE DISTORTIONS }

\author{\speaker{D.~Anchishkin}\thanks{Speaker is grateful
to the Organizing Committee for the invitation to give a talk.}\\
        Bogolyubov Institute for Theoretical Physics,
              03143 Kiev, Ukraine\\
        E-mail: \email{anch@bitp.kiev.ua}}

\author{Ye.~Anchishkin\\
        National Technical University of Ukraine "KPI",
              03056 Kiev, Ukraine \\
        E-mail: \email{anchishkin@gmail.com}}

\author{U.~Heinz\\
        Physics Department, The Ohio State University, Columbus,
              OH 43210, USA \\
        E-mail: \email{heinz@mps.ohio-state.edu}}

\abstract{
Final state interactions in two-particle interferometry is
considered for two-particle potentials which are the Coulomb plus
``strong'' one.
For a simple model of the source we provide a full numerical 
calculation of the correlation function with taking into account
two-particle Coulomb interaction plus final state strong
interaction.
It is found that for small sources ($R_0=1-4$~fm) the
contribution from the strong interaction is appreciable and
depresses the correlations while for large fireballs
($R_0=7-10$~fm) this contribution is small in comparison to the
Coulomb one. }

\FullConference{The 2nd edition of the International Workshop ---
Correlations and Fluctuations in Relativistic Nuclear Collisions --- \\
         July 7-9 2006\\
         Galileo Galilei Institute, Florence, Italy}

\begin{document}

\section{Introduction}
\label{sec1}

Two-particle correlations provide information about the space-time
structure and dynamics of the emitting source
\cite{GKW,boal,heinz99,wiedemann99,weiner99}.
Considering the correlations that occur in relativistic
heavy-ion collisions one usually assumes that:
(i)  the particles are emitted independently (or the source is completely
     chaotic), and
(ii) finite multiplicity corrections can be neglected.
Then the correlations reflect a) the effects from symmetrization
(anti-symmetrization) of the wave function and
b) the effects that are generated by the final-state interactions of the
detected particles between themselves and with the source.

The nominal quantity expressing the correlation function in terms
of experimental distributions \cite{boal} is
\begin{equation}
C({\bf k}_a,{\bf k}_b)=
\frac{\displaystyle P_2\left({\bf k}_a, {\bf k}_b\right) }
{\displaystyle P_1\left({\bf k}_a\right) \,
P_1\left({\bf k}_b\right) }
\ ,
\label{eq1}
\end{equation}
\noindent where
$P_1\left({\bf k}\right) =E\, d^3N /d^3k$
and
$P_2\left({\bf k}_a, {\bf k}_b\right) = E_a \, E_b \ d^6N /(d^3k_a\,
 d^3k_b)$
are single- and two-particle cross-sections and
$E_i=\omega(k_i)\equiv \sqrt{m^2+{\bf k}_i^2}$ is free particle energy,
$i=a,b$.

From the very first experiments while analyzing experimental data the people
wanted to separate the final state interactions (FSI) from that which is
due just to the reaction zone dynamics, as it was thought.
At first sight, the FSI can be regarded as a contamination of ``pure''
particle correlations.
However, it should be noted that the FSI depend on the structure of the
emitting source and thus provide information about source dynamics
as well \cite{anch98}.
Appreciable understanding of the particle-particle
and source-particle FSI was achieved last years
\cite{baym96,barz96,barz97,ayala98,shoppa98,sinyukov98,osada99}.
Meanwhile, there are several problems connected to the distortion of
particle-particle FSI which are not clear enough so far.
In the present paper we shall focus on two of them.
First one is an old problem of the strong final state interactions.
The second one occur in the last years due to the growth of the energy of
colliding nuclei and consequently is due to impressive increasing of the
secondary particle multiplicities
(1000 secondary pions at SPS and 2000--3000 at RHIC).

Strong FSI is a complex problem because on the fundamental level it
concerns the non-perturbative QCD.
Solution of the problem in this aspect is beyond the scope of the paper,
we solve the problem in the model approach just to have reasonable
estimations of the influence of strong two-particle interaction
(pions hard core repulsion) on the correlation function.
In different approaches it was discussed in
Refs.~\cite{suzuki87,bowler88,pratt90,anch-pr95,osada96}
(see also \cite{weiner99}, Section 3.2).

Actually, the release of pions from a small spatial region in any case is
sensitive to hard core repulsion of registered particles.
Technically this means the following:
{\bf i}. Particles created in finite volume can be represented by localized
states, for the sake of simplicity let us say by centered Gaussian packets;
{\bf ii}. To obtain correlations we integrate over
all positions of two Gaussian centers in some finite spatial volume what is
equivalent to integration over position of one center over this volume plus
integration over relative distance between centers.
The second integration includes separation distances between two particles
which are comparable with the particle (pion) core radius.
Indeed, in the pair rest system the probability $P_2$ to registrate two
particles with certain momenta ${\bf k}_a$ and ${\bf k}_b$ can be written
in the following form
(it is mostly simplified form, for details see \cite{anch98,pratt90})
 \begin{equation}
    P_2({\bf k}_a,{\bf k}_b) \approx
    \int d^3x\, \left\vert \Phi_{{\bf q}/2}({\bf x})\right\vert^2
    \int dx^0\, D(x,K)
    \, ,
 \label{i-1}
 \end{equation}
where $D(x,K)$ is the ``relative distance distribution'' of the source
expressed through the Wigner source function $S(x,K)$
($x$ is relative 4-coordinate),
$\Phi_{{\bf q}/2}({\bf x})$ is the wave function
distorted by particle-particle interaction, which describes a relative motion
of two particles, it plays a role of a ``weight function'' or probability
to find two pions separated by the distance ${\bf x}$  with relative
momentum $q=k_a-k_b$, $K=(k_a+k_b)/2$
is a mean momentum of the pair.
The integration in (\ref{i-1}) is taken over the 4-volume of the source.
If this volume is small the contribution from the region where strong
interaction dominates appreciable in comparison to contribution from
the Coulomb interaction of pions.
On the other hand, as we shall see in section II,
when the emitting volume is large one can neglect this region at all
because the contribution from strong interaction is relatively small.

The concept of freeze-out accounts for the fact that nevertheless the
particles interact always electromagnetically one can neglect this
interaction in favour of strong interaction in dense many-particle system.
Then, it is commonly accepted definition that on the background of expansion
of such a system the beginning of "freeze-out" starts just after a last
strong rescattering  and in the further evolution of the system just
electromagnetic interaction left.
Natural question can arise in this connection - is there contradiction
(double counting) with respect to freeze-out definition
if one includes strong repulsion of pions into account of FSI?
To this end let us note, if one integrates over the relative distance
in (\ref{i-1}), even from the most naive classical point
of view it is obvious, that the volume of the pion
mean radius $\langle r_\pi \rangle $ should be excluded from integration
to prevent appearance of two pions in the same domain (something
like spatial Pauli principal).
On quantum mechanical level this can be taking into account by the relevant
pion-pion repulsive potential.
Due to the strong interaction the two pions emitted from a small spatial
volume, which is comparable with pion mean radius
$\langle r_\pi \rangle \approx 0.66$~fm,
obtain always a repulsive kick which then appreciably reflected
in the correlation function, as we shall see further.
This means that for small volumes ($R\le 3--4$~fm) we cannot sharply separate
the electromagnetic and strong
interactions and must include the last pion-pion strong rescattering
in the post freeze-out consideration and by this, rigorously speaking,
we should reformulate the definition of the "freeze-out".
Actually, it is reasonable because the registered pions, in some sense,
have a {\it memory about their last strong interaction}.
Indeed, the last pion-pion strong rescattering should be accounting
separately or extracted from overall strong production amplitude
because the process of further averaging through multiple rescattering
or many-particle chaotization is no more going on after it and the last
mutual rescattering of registrated particles gives a contribution to the
post emission relatve dynamics of these particles.
It means that we separate conventionally the overall creation process
on three stages (parts):
1) fireball production,  2) strong rescattering of two registered
particles on one another, 3) particle-particle electromagnetic interaction.
This consideration is valid due to inclusive character of two-particle
spectrum (last strong interactions with other particles are integrated out).
Note, that above logics deals with quantum mechanical description
where one accounts for effects using the probability amplitude  and
two-particle wave function which is spread in space, 0thus we can
speak about the last strong interaction which gives contribution to a
certain probability even in the case when the centers of wave packets are
further from one another than $\langle r_\pi \rangle $.


\section{Final state interactions }
\label{sec2}


Two-particle probability to registrate particles with certain momenta
${\bf k}_a$ and ${\bf k}_b$
in the smoothness approximation was calculated as \cite{anch98}
 \begin{eqnarray}
&& P_2({\bf p}_a,{\bf p}_b)
=
  \int d^4x \ d^4y \ S\left(x+{\textstyle{y\over 2}},p_a\right) \,
       S\left(x-{\textstyle{y\over 2}},p_b\right)
  \bigg[ \, \theta(y^0)
  \left\vert \phi_{{\bf q}/2}({\bf y}{-}{\bf v}_b y^0) \right\vert^2
  +
\\
&&
  + \, \theta(-y^0)
  \left\vert \phi_{{\bf q}/2}({\bf y}{-}{\bf v}_a y^0) \right\vert^2 \,
  \bigg]
\pm \, \int d^4x\ d^4y\ S\left(x+{\textstyle{y\over 2}},K\right)\,
     S\left(x-{\textstyle{y\over 2}},K\right)
     \phi^*_{-{\bf q}/2}({\bf y}{-}{\bf v} y^0) \,
     \phi_{{\bf q}/2}({\bf y}{-}{\bf v} y^0)
\nonumber
\, ,
\label{50+6}
\end{eqnarray}
where $S(X,K)$ is the single particle Wigner density  of the source
 \begin{eqnarray}
  S(X,K) &=& \int d^4x\, e^{i K\cdot x}\,
  \sum_{\gamma \gamma'} \, \rho_{\gamma \gamma'}\,
  \psi_\gamma\left(X+{\textstyle{x\over 2}}\right) \,
  \psi_{\gamma'}^*\left(X-{\textstyle{x\over 2}}\right)
 \label{42}
 \end{eqnarray}
with $\rho_{\gamma \gamma'}$ as the density matrix of the source which
provides averaging over full set of singl-particle quantum numbers
$\gamma$ of the particle wave function $\psi_\gamma$ on the freeze-out
hypersurface.
We introduce pair mean momentum ${\bf K}=({\bf p}_a+{\bf p}_b)/2$ and
relative momentaum of the registrated particles
${\bf q}={\bf p}_a-{\bf p}_b$.
Using the hermiticity of the density matrix one easily shows that
$S(X,K)$ real.
Here we defined also the three velocities
 \begin{equation}
   {\bf v} = {{\bf K}\over E_K}\, ,\quad
   {\bf v}_a = {{\bf p}_a\over E_K}\, ,\quad
   {\bf v}_b = {{\bf p}_b\over E_K}
 \label{50+5}
 \end{equation}
associated with the observed particle momenta ${\bf p}_a$, ${\bf p}_b$,
and their average ${\bf K}$. The energy $E_K$ is defined as
$E_K=\sqrt{m^2+{\bf K}^2}$ and in the pair center off mass system it reduces
to particle mass $E_K=m$.

The functions $\phi_{{\bf q}/2}$ in Eq.~(\ref{50+6}) are the eigenstates
of the stationary Schr\"odinger equation
 \begin{equation}
   \hat{H}({\bf r}) \, \phi_{{\bf q}/2}({\bf r}) = E_{\rm rel}
   \phi_{{\bf q}/2}({\bf r})
 \label{16}
 \end{equation}
(where $E_{\rm rel} = {\bf q}^2/2\mu $ and ${\bf r}$ is the relative
coordinate) with asymptotic boundary conditions
 \begin{equation}
  \lim_{|{\bf r}| \to \infty} \phi_{{\bf q}/2}({\bf r}) =
  e^{\frac{i}{2} {\bf q}\cdot {\bf r}} \, .
 \label{17}
 \end{equation}
where hamiltonian $\hat{H}({\bf r})$ reads
%
 \begin{equation}
  \hat{H}({\bf r}) = - \frac{1}{2 \mu} {\bf \nabla}_r^2 + V(r)
 \label{12}
 \end{equation}
with reduced mass $\mu =m/2$ for identical particles and potential which
describes the interaction of the registrated particles after freeze-out.

In our further evaluations we will not concentrate on the dependence
of the two-particle probability $P_2({\bf p}_a,{\bf p}_b)$ on particle
velocities putting them approximately equal to the velocity
associated with mean momentum of the pair, i.e.
${\bf v}_a\approx {\bf v}$ and ${\bf v}_b\approx {\bf v}$.
Then, in pair center off mass system, where  ${\bf v}=0$, Eq.~(\ref{50+6})
reduces to the expression
\begin{eqnarray}
P_2({\bf q})
&=& \vspace{-0.5cm}
  \int d^4x_a\, d^4x_b\,
       S\left(x_a,p_a\right)\, S\left(x_b,p_b\right)\,
  \left\vert
\phi_{{\bf q}/2}({\bf x}_a -{\bf x}_b) \right\vert^2
 \nonumber\\
  && \vspace{-0.1cm} \pm
  \int d^4x_a\, d^4x_b\,
       S\left( x_a,K \right)\, S\left( x_b,K \right) \,
\phi_{{\bf q}/2}^*({\bf x}_b - {\bf x}_a) \, \phi_{{\bf q}/2}({\bf
x}_a -{\bf x}_b) \, ,
\label{50+6a}
 \end{eqnarray}
where the first term on the r.h.s. of this equation expresses the FSI
of the nonidentical particles thus it does not factorizes into the product
of the two single-particle probabilities. It has transparent physical
meaning: two single-particle probabilities to find particles in the
time-space points $x_a$ and $x_b$ with certain momenta ${\bf p}_a$ and
${\bf p}_b$, which are expressed by $S(x,p)$, is weighted by the probability
$\left\vert \phi_{{\bf q}/2}({\bf x}_a -{\bf x}_b) \right\vert^2$
to find these particles with relative distance ${\bf x}_a -{\bf x}_b$ and
relative momentum ${\bf q}$. The correlation function is the ratio of the
obtained probability $P_2({\bf q})$ Eq.~(\ref{50+6a}) to the product of
two single-particle probabilities, it reads
 \begin{equation}
  C({\bf q}) =
  \frac{\displaystyle P_2({\bf q}) }
{\displaystyle   \int d^4x_a\, S\left(x_a,p_a\right)\,
                 \int d^4x_b\, S\left(x_b,p_b\right) }
\, ,
 \label{2-11}
 \end{equation}
where 4-vectors $p_a=({\bf q}^2/4m,\, {\bf q}/2)$ and
$p_b=({\bf q}^2/4m,\, -{\bf q}/2)$.

In the noninteracting limit
$\phi_{{\bf q}/2}({\bf x}) \to \exp{(i{\bf q}\cdot {\bf x}/2)}$ one
recovers known expression for the correlation function
 \begin{equation}
  C({\bf q},{\bf K}) = 1
  \pm
  \frac{\displaystyle  \left\vert  \int d^4x\,
 e^{i{\bf q}\cdot {\bf x}}   S\left( x,K \right)\, \right\vert ^2}
{\displaystyle   \int d^4x_a\, S\left(x_a,p_a\right)\,
\int d^4x_b\, S\left(x_b,p_b\right) }
\, .
 \label{2-12}
 \end{equation}
%

\subsection{Application. Coulomb plus strong final state interaction}
\label{sec2-2}

To make evaluations of the correlation function (\ref{2-11}) we take
the source function in the form
\begin{equation}
S(x,p) =
\frac{e^{-p^0/T_{\rm f}}}{4\pi m^2TK_1(m/T_{\rm f})}\,
\frac{e^{-x_0^2/2\tau^2}}{(2\pi)^{1/2}\tau}\,
\frac{e^{-{\bf x}^2/2R_0^2}}{(2\pi)^{3/2}R_0^3}
\, .
\label{4-1}
\end{equation}
Then, in the pair c.m.s (${\bf K}=0$) the expression for two-particle
probability (\ref{50+6a}) reduces to the form
\begin{eqnarray}
P_2({\bf q})
=
\frac{e^{-2K^0/T_{\rm f}}}
          {[4\pi m^2TK_1(m/T_{\rm f})]^2\, \pi^{3/2}\, (2R_0)^3} \,
   \bigg[ &&  \int d^3r \, e^{-r^2/4R_0^2}
   \left\vert \phi_{{\bf q}/2}({\bf r}) \right\vert^2
\nonumber \\
&&
  \pm
   \int d^3r \, e^{-r^2/4R_0^2} \,
   \phi_{{\bf q}/2}^*(-{\bf r}) \,
   \phi_{{\bf q}/2}({\bf r})   \bigg]
\, ,
\label{4-2}
\end{eqnarray}
where time dependence is integrated out.
In the present simple model of $S(x,p)$ (\ref{4-1}) the single-particle
probability to registrate particle with definite momentum ${\bf k}$ reduces
to the pure Boltzmann exponent, i.e.
$P_1({\bf k})=[4\pi m^2TK_1(m/T_{\rm f})]^{-1}
 \exp{[-\omega({\bf k})/T_{\rm f}]}$,
where time and spatial dependencies are integrated out.
Thus, we can write now the correlation function (\ref{eq1}) in the form
 \begin{equation}
  C({\bf q}) \ = \
\frac{1}{\pi^{3/2}\, (2R_0)^3} \,
   \int d^3r \, e^{-r^2/4R_0^2}
   \left[ \,  \left\vert \, \phi_{{\bf q}/2}({\bf r}) \, \right\vert^2 \,
  \pm \,
   \phi_{{\bf q}/2}^*(-{\bf r}) \,
   \phi_{{\bf q}/2}({\bf r})\,   \right]
\, ,
 \label{4-3}
 \end{equation}
where the Boltzmann exponents canceled because
$2K^0=\omega({\bf k}_a)+\omega({\bf k}_b)$ and we adopt that the Boltzmann
factors have the same freeze-out temperature $T_{\rm f}$
for the two-particle and single-particle spectra
(the single-particle spectra are taken usually from different
collision event than the two-particle spectrum).

The numerical evaluations of the distorted wave function
$\phi_{{\bf q}/2}({\bf r})$
is provided by solution of the Schr\"{o}dinger equation
with the relevant potential which reflects two-pion interaction.
We solved this equation by partial expansion of the wave function
for the Coulomb potential energy
$V_{\rm Coul}(r)=\alpha/r$ and for the potential energy which accounts
for Coulomb interaction plus strong interaction of the frozen pion pair
$V_{eff}(r)=V_{\rm Coul}(r)+V_{\rm str}(r)$,
where ``strong potential'' reads
$V_{\rm str}(r)=V_0 \exp{(-m_\rho r)}/(m_\rho r)$ with
$V_0=2.6$~GeV, $m_\rho =770$~MeV.
Potential of the strong repulsion \cite{pratt90} was chosen to match
the behavior of the pion-pion scattering phase shifts.
In any case the use of this strong potential can be considered as a model
of the short-range repulsion which is possessed by pions.
The correlations functions which correspond to these potentials are plotted
in Figs.~1-4 for different samples of the freeze-out fireball radii:
$R_0=2,\, 4.5,\, 7.1,\, 10$~fm. It is evidently seen in Figures that
the contribution of the strong potential is appreciable for small radii
emission sources, i.e $R_0=2$~fm and $R_0=4.5$~fm.
For the source size which is of the
$^{207}$Pb radius $R_0=7.1$~fm (see Fig.~3)
the contribution which can be put in correspondence to the strong potential
is comparatively small.
And for the source of larger radius, for instance $R_0=10$~fm,
the influence of the strong FSI even negligible (see Fig.~4).

Remind, that  in the absence of the two-particle FSI the source function
(\ref{4-1}) results in the correlation function (${\bf K}=0$)
$C({\bf q}) = 1 \pm \exp{(-{\bf q}^2R_0^2)}$.
For completeness of comparison we plot in Figures this correlation
function, corrected by the Gamov factor $G(|{\bf q}|)$,
which for boson-boson correlations takes form
 \begin{equation}
  C({\bf q}) \ = \
  G(|{\bf q}|) \left( 1 \, + \, e^{-{\bf q}^2R_0^2} \right)
\, ,
 \label{4-5}
 \end{equation}
where
\begin{equation}
G(|{\bf q}|)=
\mid \phi_{{\bf q}/2} ({\bf r}=0)\mid ^2= \frac{2\pi \eta }{e^{2\pi \eta }-1}
\label{4-6}
\end{equation}
with $\eta = \alpha m_{\pi }/|{\bf q}|$.
It is seen in Figures 2, 3 and 4 that the finite
size of the emission source  softens the manifestation of the FSI and
the `Gamov factor' tends to overestimate the FSI effects for the source
of big size ($R_0\ge 4$~fm).

As it seen in Figs.~1, 2 additional repulsive potential (hadron hard core)
essentially suppressed the correlation function at small relative momentum.
Actually, the physical reasons for this are transparent:
two strongly interacting particles emitted from the volume of
$R_0\approx 2--3$~fm,  when their "own mean radius", reflected by repulsive
strong potential $V_{\rm str}$, is about
$\langle r \rangle \approx 0.66$~fm, should sufficiently "feel" one another
through mutual repulsion.
On the other hand, when emision zone is much larger than the particle
mean radius
$\langle r \rangle $, for instance fireball radius is $R_0=10$~fm,
the contribution of the short-range repulsion is negligible as it seen
in Fig.~4.
As a matter of fact, the last can be regarded as undoubt indication
that correlations of the particles which are emitted from the large
separate distances, starting for instance from $R\ge 4$~fm,
are not practically influenced by hard strong repulsion,
or in other words, for the particles which are emitted from perepherical
regions the contribution from short-range interactions is washed out and
their correlations are not suppressed by strong interaction.

It is very interesting to put this conclusion in connection with the so
called concept of the  "dominant contribution to the correlation function of
the regions of spatial homogeneity".
This notion states that on the scale of the collective
(hydrodynamical) velocities of two local domains it is difficult (small
probability) to find two particles with approximately equal momenta.
(For instance, the most noticeable example is two local parts of the
fireball which move in the opposite directions to one another.)
Thus, the contribution to the correlation function from the regions
which spatially stand far from one another is suppressed due to collective
motion.
Actually, in the present simple model we do not consider expansion of
the source.

%
\begin{figure}
\includegraphics[width=.6\textwidth]{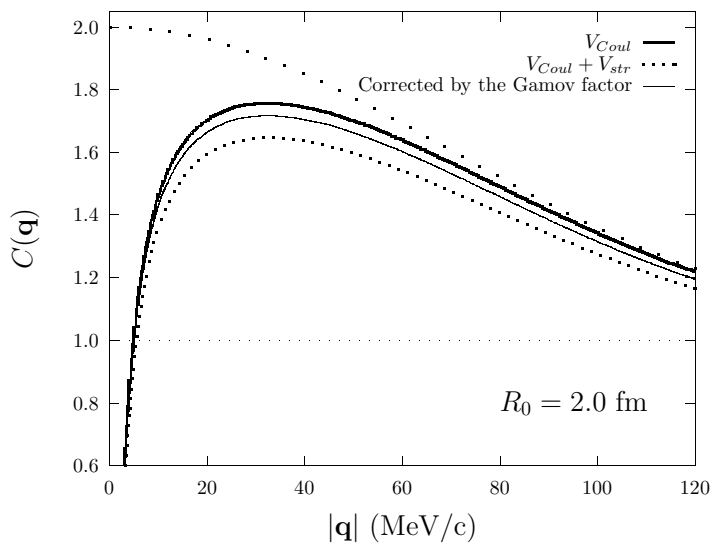}
\caption{Correlation function versus relative pion momentum,
$R_0=2.0$ fm.
Dotted (far-between) curve is the correlation function without FSI.
Fat solid curve corresponds to the two-particle Coulomb potential
$U_{Coul}(r)=\alpha/r$.
Dotted curve corresponds to the sum of the Coulomb and strong
two-particle potentials
$U(r)=\alpha/r+V_0 \exp{(-m_\rho r)}/(m_\rho r)$.
The bottom solid curve is the correlation function corrected just by
the Gamov factor.}
\label{fig1}
\end{figure}

On the other hand, as we have just proved
a short-range repulsion of the strongly interacting
particles emitted from a small domain ($R\le 4$~fm) decreases as well
the probability to find two particles with a small relative momentum,
for example if the volume of homogeneity is of the size $R\approx 2-4$~fm,
then the correlation function of pions from this region,
as we see from Figs.~1, 2,
will be suppressed by strong repulsion of these particles.
Hence, both effects work in the same
direction, decreasing the correlations of the particles emitted from the
small regions (hard core repulsion) and decreasing the correlations of
the particles emitted from the far separated regions (collective motion).
If these effects have the same scale then they approximately uniformly
decrease correlations of two particles which are emitted from all regions
of the fireball.
%
\begin{figure}
\includegraphics[width=.6\textwidth]{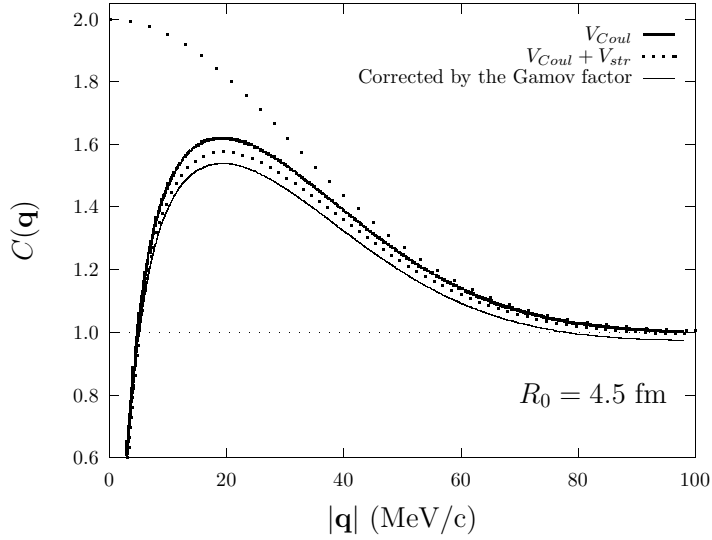}
\caption{ The same as in Fig.~1, but for $R_0=4.5$ fm. }
\label{fig2}
\end{figure}
%
The last statement results in conclusion that two-particle correlations
does not occur preferably from the small regions where the spatial separation
of the particles is small (hence, they have the same collective velocity),
but the particles which are emitted from two regions which are separated by
large distance contribute also appreciably to the correlation function.
Thus, the correlation function shows us the real size of the fireball
at freeze-out, but not just a size of the small region of homogeneity.
If the region of spatial homogeneity has the radius
$R\ge 7$~fm then the strong repulsion can be neglected.
So, the strong FSI effectively extends the size of the region from which
the particles are allowed for correlations.
If we unite two cases where the radius of homogeneity is small
$R_{\rm homo}\le 4$~fm and large $R_{\rm homo}\ge 7$~fm (hence, it is comparable with
the size of the fireball)
then we come to the conclusion that the radius which is
extracted from the correlation function is approximately the radius of the
fireball.



%
\begin{figure}
\includegraphics[width=.6\textwidth]{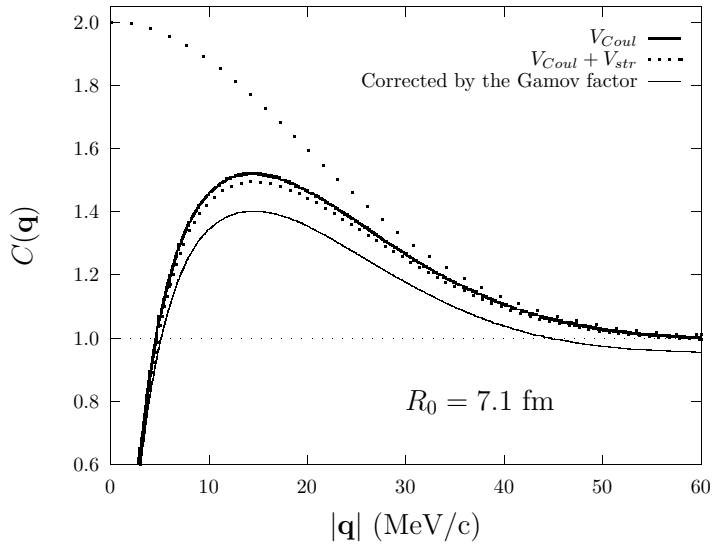}
\caption{ The same as in Fig.~1, but for $R_0=7.1$ fm. }
\label{fig3}
\end{figure}

%
\begin{figure}
\includegraphics[width=.6\textwidth]{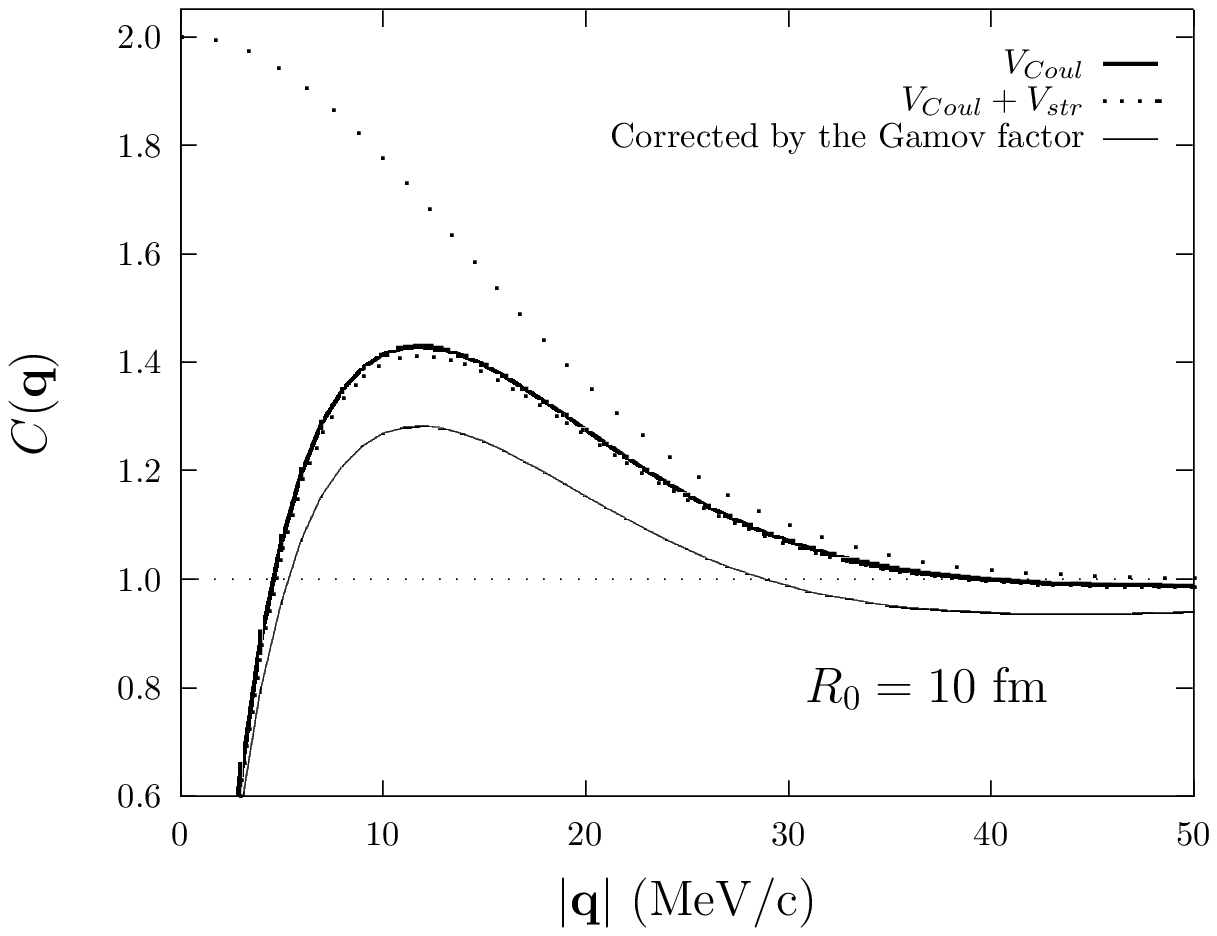}
\caption{ The same as in Fig.~1, but for $R_0=10$ fm. }
\label{fig4}
\end{figure}

It is interesting to point out that for the sources of the size
$R_0\approx 3$~fm correction of the correlation function just by the
Gamov factor effectively gives a right correction, literally it becomes
equal to the account of a finite size of the source exploiting potential
$V_{eff}(r)$ which includes strong FSI. It means, that if one starts from the
Gamov factor correction then account for finite size of the source softens
Gamov's correction. If we include strong interaction (in addition to Coulomb
one) then for the source
of the small size account for finite size of the source makes
correction more hard
(in comparison to Gamov factor correction). But in the intermediate point,
which is $R_0\approx 3$~fm these two deviations from the Gamov corrections
cancel one another


\section{Summary and conclusions}
\label{sec4                                             }


We examined pion-pion final state interaction when the two-particle
potential energy is not just the Coulomb one.
Distortion of the Coulomb potential is taken in the additive
form, what is due to account for particle-particle strong interaction.
To make our investigation transparent as much as possible we take: \\
a) for the description of the fireball (particle emitting source),
   the Wigner source function $S(x,K)$ in the Gaussian form;
b) small relative momentum approximation what results in equlity of two
   particle velocities and pair c.m.s. velocity, then expression for the
   two-particle probability is sufficiently simplified,
   see Eq.~(\ref{50+6a}).

We argue that for inclusive two-particle spectrum one should explicitly
account for strong interaction of two registered particles because their
mutual last rescattering essentially determines dynamics
of their post-freeze-out relative motion.
Strong two-particle potential was chosen as screened Yukawa potential and
reflects short-core repulsion of pions.
Then the wave function describing the relative motion of pions was founded
as numerical solution of the Schr\"odinger equation where potential is
taken as sum of the Coulomb and Yukawa potentials.
We obtained that correlation function is sensitive to the presence of short
range two-particle interaction when emission volume is of the radius
$R_0\le 4$~fm, see Figs.~1 and 2.
Indeed, two strongly interacting pions emitted from the separate distance
(separation of the centers of the wave packets),
which is comparable to the pion "mean radius"
$\langle r_\pi \rangle \approx 0.66$~fm,
obtain a mutual repulsive kick which is then reflected
as decreasing of the two-particle probability to find particles with
"initial production" relative momentum, i.e. the "strong kick" shifts
along x-axis this particular probability value to higher values of the
relative momentum.
Consequently, this will result in suppression of the correlation function what
is most noticeable for small relative momentum.
(Obviously, the effect which is due to the strong repulsion is additional
to that one which is caused by the Coulomb repulsion.)

On the other hand the presence of strong particle-particle repulsion almost
unnoticeable in the correlation function when one considers the sources of
big size, $R_0\ge 7$~fm, see Figs.~3 and 4.
This just confirms that the hard core repulsion of the registered
particles
appreciably influences the correlation function when the particles
are emitted from the region of a small size.
If the region of homogeneity is also small $R_{\rm homo}\le 4$~fm
then the correlations originated by pions emitted from this region
are suppressed.
This suppression can be of the same strength as suppression of the
correlations of pions emitted from the domains which are beyond the radius
of homogeneity (the latter suppression is due to collective motion of
expanding system) what results in uniform suppression of correlations of
two pions emitted from
Then, the radius $R_0$  which is measured by the correlation function can
be larger than the size of homogeneity region and tends to cover the
overall size of the fireball.
In the opposite case, when the size of the region of
homogeneity is larger $R_{\rm homo}\ge 7$~fm, the correlation function
is not influenced practically by the strong interaction of two registered
pions.
Hence, the radius $R_0$ which is extracted from the correlation function
in this case is about $R_{\rm homo}$ and
$R_0$  really reflects the size of homogeneity volume.

Summarizing all above we can say: \\
\noindent
The strong final state interactions can be neglected when the volume of
spatial homogeneity in the fireball are of the size $R_{\rm homo}\ge 7$~fm.
Then, the radius $R_0$ which is extracted from the pion-pion correlation
function measures the size of the regions of homogeneity in the emitting
source and coincide with $R_{\rm homo}$.
When the radius of homogeneity volume is small $R_{\rm homo}\le 4$~fm
the strong final state interactions are noticeable and their presence
is reflected in the correlation function in the way that radius $R_0$
(extracted from the correlation function) is larger than $R_{\rm homo}$
and tends to be the overall fireball radius. \\

\bigskip

{\bf Acknowledgements:}
The work of D.A. was supported by the Regensburg University under
BMBF Project 06 OR 883.
D.A. wishes also to express his deep appreciation to the staff of CERN TH
Division for warm hospitality.
The work of U.H. was supported by the U.S.
Department of Energy under contract DE-FG02-01ER41190.

\end{document}